\documentclass[conference]{IEEEtran}

\usepackage{cite}
\usepackage{amsmath,amssymb,amsfonts}
\usepackage{algorithmic}
\usepackage{graphicx}
\usepackage{textcomp}
\usepackage{xcolor}

\usepackage{multirow}

\def\BibTeX{{\rm B\kern-.05em{\sc i\kern-.025em b}\kern-.08em
    T\kern-.1667em\lower.7ex\hbox{E}\kern-.125emX}}

\usepackage[utf8]{inputenc} 
\usepackage[T1]{fontenc}    
\usepackage{hyperref}       
\usepackage{url}            
\usepackage{booktabs}       
\usepackage{nicefrac}       
\usepackage{microtype}      
\usepackage{lipsum}
\usepackage{diagbox}
\usepackage{caption}
\graphicspath{ {./images/} }

\begin{document}

\title{SemChunk-C: Semantic Segmentation for C Code}


\author{
\IEEEauthorblockN{Boris Nazarov, Darya Frolova, Shaked Leibzirer,  Pavel Kisilev}

\IEEEauthorblockA{
\texttt{\{boris.nazarov, darya.frolova, shaked.leibzirer, pavel.kisilev\}@huawei.com}
}}





\maketitle

\begin{abstract}

Semantic segmentation of code written in a C-family language remains a challenging problem, due to the language's complex syntax, macro expansion, and irregular structural patterns. Existing chunking methods, such as fixed-sized windows, heuristic splitting, and syntax-based tools, often fail to capture meaningful functional units, limiting the efficacy of  retrieval and other downstream LLM driven tasks.

In this paper, we address the problem of chunking in C-related languages. First, we define a set of code chunk categories. Second, we train an LLM-based classifier to a) identify chunk boundaries, and b) assign each chunk a descriptive functional attribute (a category), which can be useful for downstream tasks. By leveraging the LLM’s ability to capture semantic context within the code, we assume flexible chunk boundaries, allowing to adapt to the specific structure and context of each instance. Third, we introduce SemChunk-C, a family of lightweight language models for semantic chunking of C-related files (.c, .cpp, .h, .cs, etc.).
These models are based on the first four Ettin encoders \cite{ettin} with 17M, 32M, 68M, and 150M parameters. Despite their relatively small size, they are capable of identifying cohesive code units, such as data structures, interface blocks, and other components. Furthermore, we demonstrate the robustness of our approach on real-world code, including challenging constructs such as nested definitions and macros.

 We test our approach on various datasets, and show that it  achieves high boundary accuracy and semantic coherence, matching or outperforming chunkers that are based on much larger code-oriented LLMs. We also validate the improved performance of the downstream tasks on a few curated benchmarks.

\end{abstract}


\section{Introduction} 
\label{sec:introduction}

Recently, there has been a growing tendency to employ small language models (SLMs) as encoders and embedders.  Such small models are often sufficient, since representation learning tasks require capturing semantic similarities rather than performing complex reasoning or text/code generation. Compared to large generative models, SLMs offer significant advantages in terms of computational efficiency, latency, and deployment cost. Furthermore, they are able to produce high-quality text representations when properly trained for domain-specific data. In addition, the use of SLMs for encoders and embedders provides important privacy advantages. Their reduced computational and memory requirements make it feasible to deploy them locally on edge devices, minimizing the need to transmit sensitive data to external servers. This local processing capability reduces exposure risks and enables privacy-preserving applications in many domains. Thus, small models not only offer efficiency benefits, but also play a critical role in building secure and trustworthy semantic systems.

Equally important is the role of semantic chunking in representation-based systems. Rather than dividing text into fixed-length segments, semantic chunking partitions content according to its meaning, discourse structure, or topical coherence. This approach improves embedding quality by ensuring that each chunk represents a self-contained semantic unit. As a result, semantic chunking enhances retrieval accuracy and downstream performance in tasks such as search, retrieval-augmented generation (RAG), and document understanding.

Our goal is twofold — use small language models, and create \textit{semantic} chunking. We demonstrate that SLMs can effectively segment text, with a specific focus on the relatively underexplored task of \textit{code chunking}. We show that, despite the challenges of this specific domain, small models can produce high-quality and semantically meaningful code segmentation.

As the use case, we chose C-family languages. Large codebases written in C, C++ and C\# remain difficult to process using language-model-based systems due to their complex syntax, heavy use of macros, and weak structural consistency. Existing chunking strategies, such as fixed-size token windows, heuristic splitting, or syntax-based tools such as Tree-sitter, often fail to align with actual semantic units of code. These alignment problems may lead to poor retrieval performance, and to inefficient context construction for downstream LLM tasks.
In this work, we introduce SemChunk-C - a family of lightweight language models capable of performing semantic chunking of C-family files (.c, .cpp, .cs, .h, .m/.mm). The models a) identify meaningful code chunks (e.g. cohesive functions, interface blocks, data structure definitions and other components), and b) assign each chunk a descriptive attribute that captures its functional role. Despite the modest model sizes (17M - 150M parameters), they demonstrate strong semantic understanding and robustness to real-world complexities, such as nested declarations, macros, and template-heavy C and C++ code.
We evaluate our approach against rule-based chunker and also the use of larger language models. Across a curated benchmark of C/C++ source files, our method achieves higher chunking accuracy and produces more coherent and functionally homogeneous chunks. Also, our approach yields substantial improvement in downstream tasks, such as code retrieval and generation. 
One of the key advantages of our token classification approach is that it helps avoid hallucinations, which is a key limitation of purely LLM-based chunking. Large models may introduce spurious content, such as additional code, extraneous text, or unwanted attributes, which we have observed in practice.

Our results suggest that small domain-specialized LMs can outperform significantly larger models on targeted semantic segmentation tasks, offering a practical and efficient alternative for code indexing, repository summarization, and intelligent developer toolkit.

Although the chunking process for large existing repositories may appear time-consuming, the advantages of employing semantic chunks for repository-based downstream tasks are substantial. Therefore, incorporating semantic chunking into LLM-based code generation workflows can be highly beneficial.

The remainder of this paper is organized as follows: we first review related work, then present our method, and finally describe the experimental setup and results.
\section{Related Work}
\label{sec:related_works}

The efficacy of modern code-intelligence tasks, ranging from RAG and automated code completion to repository-level understanding, is fundamentally governed by how source code is partitioned. As context windows remain a finite resource, the transition from heuristic-based splitting to semantic chunking has become an important software engineering research.

\subsection{Foundation of Semantic Chunking}
The challenge of dividing a continuous stream of text into meaningful segments has its roots in classical natural language processing. Early foundational work, such as the \textbf{TextTiling} algorithm proposed in \cite{hearst1997texttiling}, utilized lexical cohesion and multi-paragraph patterns to identify subtopic boundaries. As the field shifted toward vector space models, research transitioned to using semantic word embeddings to detect shifts in coherence \cite{alemi2015text}, while subsequent supervised learning approaches \cite{koshorek2018text} laid the groundwork for modern "semantic splitting."

In the current era of LLMs, the focus has moved toward identifying the optimal "retrieval unit." Recent work by\cite{gao2023dense} introduces the concept of \textbf{Propositional Chunking}, arguing that retrieving atomic "propositions"—sentences distilled into independent factual units—significantly outperforms traditional paragraph-based retrieval. This evolution is driven by the realization that chunking strategy is a primary "failure point" in production-level RAG systems \cite{barnett2024seven}. By calculating embedding similarity across these units, modern systems can establish boundaries where the thematic divergence exceeds a statistical threshold, such as an interquartile range or standard deviation shift, ensuring that each chunk captures a complete, independent idea.

\subsection{Code-Specific Structural Challenges}
Unlike natural language, source code is best represented as a structured semantic graph rather than a linear narrative. Effective chunking must take into account programming constructs such as functions, classes, and variable scopes, in order to avoid breaking logical dependencies. The foundational work in this area established that capturing intent requires moving beyond raw text to \textbf{identifier-based embeddings} \cite{efstathiou2019semantic} and deep semantic parsing \cite{graphcodebert2020}. 

Recent literature has focused on bridging the gap between syntax and semantics through Abstract Syntax Trees (ASTs). For example, \textbf{CodeRAG} \cite{zhang2025coderag} and \textbf{CodeWisp} \cite{codewisp2025} utilize AST-guided structures to provide highly relevant context for repository-level completion. Similarly, \textbf{cAST} \cite{cast2025} uses recursive parsing methods to merge semantically linked blocks while respecting strict token limits. However, a significant limitation persists: these syntax-reliant methods depend on strict syntactic validity. In C-family languages, the presence of complex macro expansions often causes standard parsers to produce fragmented or failed trees, leaving "functional atoms" obscured.

\subsection{Model-Driven Segmentation and Efficiency}
To overcome the fragility of ASTs, a new wave of "model-first" chunking has emerged. Frameworks like \textbf{CODE2JSON} \cite{code2json2025} and \textbf{RANGER} \cite{ranger2025} utilize zero-shot LLM agents to extract natural language representations and knowledge graphs directly from code. The value of this semantic approach is evident in downstream tasks; \textbf{FuncVul} \cite{halder2025funcvul} demonstrates that precise function-level chunking is critical for vulnerability detection, while \textbf{LongCodeZip} \cite{longcodezip2025} employs a two-stage strategy to compress and chunk long-context repositories.

Furthermore, empirical evidence suggests that models may now surpass human performance in this domain. Recent investigations by \cite{glasz2025llm} found that LLM-generated partitions yield summary comments that are up to 20\% more factual than human-created boundaries. However, this superior performance comes at a prohibitive computational cost. Relying on massive LLMs for initial indexing of large-scale C-family repositories is not feasible for real-world deployment. 

Our work addresses this gap by introducing a family of lightweight chunking models based on \textbf{Ettin encoders} \cite{ettin} (17M to 150M parameters) that identify flexible semantic boundaries within irregular patterns of C-family code, achieving LLM-level precision with the throughput required for scalable software engineering.

\section{Data, Models and Training}
\label{sec:our_method} 

We investigate the potential use of small language models for code understanding related tasks. We focus on a subset of the Ettin family of encoder models with 17M, 32M, 68M, and 150M parameters \cite{ettin}. These models share an identical architecture, which allows them to be fine-tuned using the same training recipe. This, in turn, allows us to investigate in a systematic way a smallest model which is still capable of good chunking performance.

\begin{figure}
        \centering        \includegraphics[width=1\linewidth]{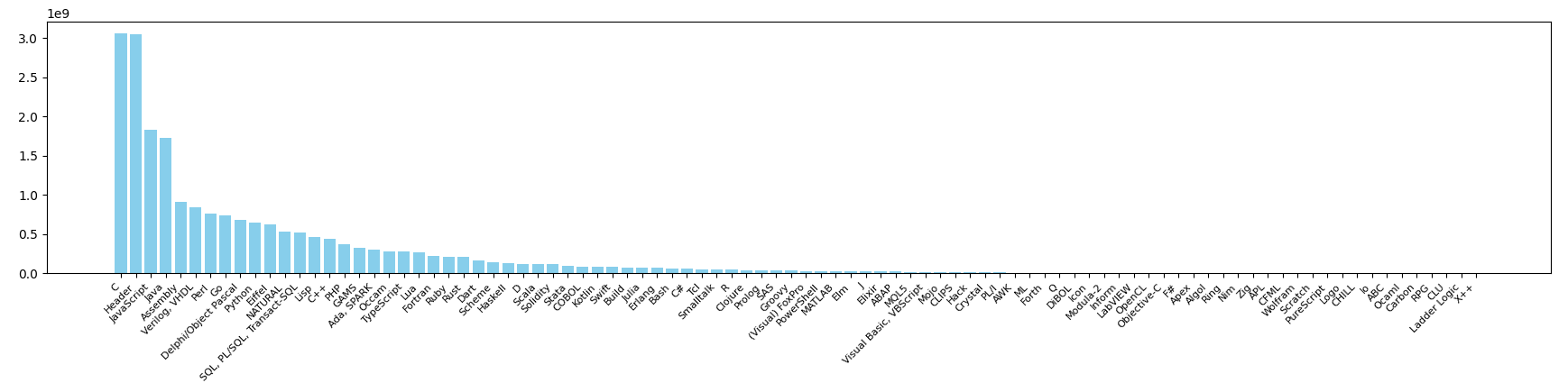}
        \caption{Downloaded languages distribution used for training of our small language encoder-only models.}
        \label{fig:PLs}
\end{figure}

First, we fine-tuned the models in an unsupervised manner on publicly available repositories covering multiple programming languages (PLs). Then, using a Qwen2.5-Coder-32B-Instruct LLM  we automatically generated a training dataset consisting of (chunk, attribute) pairs specific to C-based languages. Finally, fine-tuned Ettin encoder models of various sizes were trained for token classification to segment files into chunks, and to assign an attribute to each chunk. More details are provided below.

\subsection{More Code Knowledge Means Better Performance}
\label{subsec:FT}

As the base model for our chunker, we train a small encoder model with strong capabilities in code syntax. Although the selected Ettin models were already trained on more than 20B tokens of code data, we further enhanced them with additional code, placing particular emphasis on C-related programming languages.

We downloaded nearly 27B tokens from multiple repositories, containing more than 90 PLs, prioritizing C-related languages. Other PLs were used as regularization and augmentation data to enhance robustness and generalization. Figure \ref{fig:PLs} illustrates the distribution of the downloaded PLs.

Training hyperparameters were selected as in Ettin recipe, with masking ratio of 15\%.
To partially mitigate the effects of catastrophic forgetting, during the training we included general, non-code data which is a part of C4 dataset \cite{c4} containing nearly 5B tokens.

To evaluate the model's learned ability to better understand C code structure, we compared the base (BS) encoders with the fine-tuned (FT) ones on an internal masked-completion dataset. This dataset was constructed from 100 C language source files in which the function types were masked. Table \ref{tab:function-type-masking} presents a comparison of the base and fine-tuned models in terms of top-1, top-5 and top-25 masked-completion accuracy. As shown, the completion accuracy for the code-specific entity (function type) improves across all the models after fine-tuning. The greatest gain is observed for the smallest model (17M), which achieves a 17.35\% increase in the top-1 accuracy.

\begin{table}[]
\centering
\setlength{\tabcolsep}{4pt}
\footnotesize
\begin{tabular}{|l|c|c|c|c|}
\hline
  & 17M & 32M & 68M & 150M  \\
\hline
 &
 \begin{tabular}{c|c} BS & FT \end{tabular} &
 \begin{tabular}{c|c} BS & FT \end{tabular} &
 \begin{tabular}{c|c} BS & FT \end{tabular} &
 \begin{tabular}{c|c} BS & FT \end{tabular} \\
 \hline
top-1 & 
\begin{tabular}{c|c} 58.9 & \textbf{76.3} \end{tabular}& 
\begin{tabular}{c|c} 72.5 & \textbf{84.9} \end{tabular}&	
\begin{tabular}{c|c} 85.4 & \textbf{92.8} \end{tabular}& 
\begin{tabular}{c|c} 92.2 & \textbf{93.4} \end{tabular}\\
\hline
top-5 & 
\begin{tabular}{c|c} 79.6 & \textbf{91.0} \end{tabular}& 
\begin{tabular}{c|c} 88.7 & \textbf{97.0} \end{tabular}& 
\begin{tabular}{c|c} 94.3 & \textbf{97.5} \end{tabular}& 
\begin{tabular}{c|c} 97.7 & \textbf{98.0} \end{tabular}\\
\hline
top-25  & 
\begin{tabular}{c|c} 88.5 & \textbf{95.0} \end{tabular}&	
\begin{tabular}{c|c} 92.9 & \textbf{98.6} \end{tabular}& 
\begin{tabular}{c|c} 97.0 & \textbf{98.9} \end{tabular}& 
\begin{tabular}{c|c} 99.0 & \textbf{99.4} \end{tabular}\\
\hline
\end{tabular}
\caption{A comparison of the mask completion task accuracy of the two encoder-only models: the base (BS) model and our fine-tuned (FT) variants of different model sizes.}
\label{tab:function-type-masking}
\end{table}

After our encoders are trained to gain a better understanding of various C-code syntax, we transform them into chunker models.
In the following sections, we describe the chunker training procedure and the test sets, and further demonstrate the benefits of the fine-tuning stage for the trained chunker model (see Sec.\ref{subsec:justify-pretrain}).

\subsection{Taxonomy of Semantic Chunk Types}
\label{subsec:chunks-taxonomy}

Our chunking scheme follows a simple principle: each chunk corresponds to a self-contained, semantically meaningful unit of code. To ensure that our chunker operates at the semantic level rather than as a purely rule-based text splitter, we incorporate additional constraints that require analysis and understanding of the code. These include, for example, handling nested structures and grouping semantically related elements into coherent units. 

\begin{table}[]
\centering
\setlength{\tabcolsep}{3pt}
\begin{tabular}{cc}
    \scriptsize
    \begin{tabular}{|c|c|c|}
    \hline
      Language & Files & Tokens  \\
      \hline
    C (c) &	10,799  & 33,826,865 \\
    C++ (cpp, cc, cxx) &	7560  & 9,818,549 \\
    C\# (cs) &	10,160 & 9,266,131 \\
    H (h, hpp, hh, hxx) &	7,814  & 14,264,406\\
    M (m) &	2,277  & 4,808,555 \\
    MM (mm) &	761  & 1,996,592 \\
     \hline
    \end{tabular} &
    \scriptsize
    \begin{tabular}{|c|c|}
    \hline
    Category & Chunks\\
    \hline
      comment & 26,346 \\
      function & 120,538 \\
      class & 20,247 \\
      directive & 31,967\\
      include & 40,319 \\
      type definition & 15,910\\
      variable & 10,196\\
      namespace & 15,910\\
      interface+ & 5,839\\
      misc & 4,212\\
     \hline
    \end{tabular}
\\
\end{tabular}
\caption{The left table depicts the number of files and tokens per programming language used for the training of our chunkers. The right table depicts the distribution of chunk categories. Here 'interface+' category corresponds to 'interface', 'protocol' and 'implementation' related chunks, altogether.}
\label{tab:train-data}
\end{table}

When building a cross-language repo analyzer it is useful to create a clear structure of semantic chunk types. Below we roughly split semantic chunks into groups,  aiming to combine the following types of chunks: C-specific, Object-oriented C++, C\#-specific, and common C-family. We also add to each group an attribute (category) that describes its functional purpose. Our goal is to achieve a balance between the granularity of semantic meaning and the number of possible attributes, ensuring that a small model can associate each code snippet with the most appropriate description. We believe that organizing the code into these categories improves the formation of logically and semantically consistent structures. This organization also facilitates repository-level operations, including the systematic addition and removal of code components. Furthermore, it supports the identification of missing or superfluous dependencies. And, finally, it enhances the ability to analyze and answer questions regarding the code’s overall structure and functionality.

The following presents the list of selected chunk categories.

\begin{figure}
    \centering
    \includegraphics[width=1\linewidth]{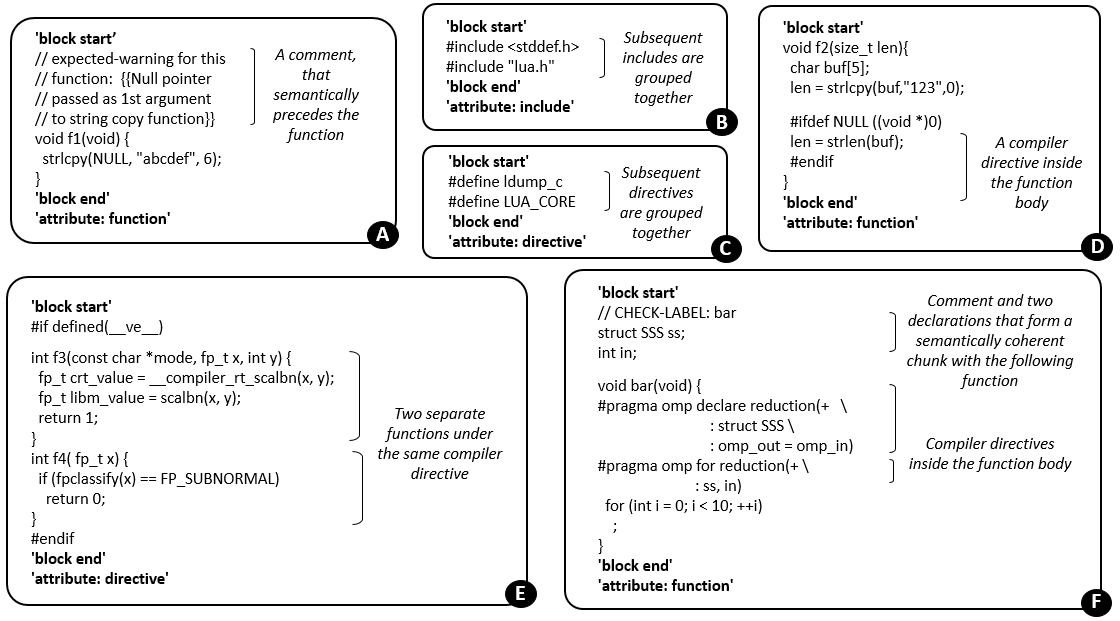}
    \caption{Examples of semantic chunking for C code.}
    \label{fig:chunks-examples}
\end{figure}

\begin{itemize}
    \item Category: \textit{'comment'}. Comments may be a single line //.., block comments /*..*/, or doxygen/documentation comment: ///, /**..*/. Comments can be standalone or attached to a declaration/implementation, so they should be chunked differently. We expect standalone comments to form a separate block, while comments that, e.g. belong to a function, should not be separated from the function body.
    See examples in Fig. \ref{fig:chunks-examples} A and F.
    
    \item Category \textit{'function'} contains function/method declarations, definitions, implementations and modules. A function may be a free function, member or inline function, template, lambda definition, operator overloads, record, orextern for functions. If a function contains sub-functions or compiler directives inside, we assume the chunker still defines it as a single 'function' block. See for example Fig. \ref{fig:chunks-examples} A,D,F.
    \item Category \textit{'class'} covers the code related to classes, methods, constructors and destructors.
    \item Category \textit{'directive'}. This category contains pre-processor and compiler directives: \#define, \#undef,  \#if / \#ifdef / \#ifndef / \#else / \#elif / \#endif, \#pragma, \#error etc.
    \item Category \textit{'include'}. An include may be one of the following types: \#include, using, import, @import or export.
    \item Category \textit{'type definition'} covers structs, unions, constant definitions, enums, declarations, typedefs and alias.
    \item Category \textit{'namespace'}. Even lengthy code segments that contain heterogeneous categories within namespace scopes should be treated as a single chunk and assigned the category \textit{namespace}.
    \item Categories \textit{'protocol'},  \textit{'implementation'}, \textit{'interface'}. These categories are relatively straightforward to detect, as the corresponding code begins with well-defined identifiers and has clearly delineated boundaries.
    \item Category \textit{'variable'}. If the code defines global or local variables, delegate declarations, or external variable declarations, these are treated accordingly. Variables declared within functions, structs, or other constructs are not separated from their enclosing implementation bodies. Furthermore, certain variables defined outside functional code blocks are associated with the corresponding code segments, as this better reflects their semantic relationships and is expected to improve performance in downstream tasks. See Fig. \ref{fig:chunks-examples} E for example.
    \item Category \textit{'misc'}. If a code snippet does not correspond to any of the types, described above, the attribute 'misc' is added.
\end{itemize}

\textit{Nested and interleaved structures.} Some categories, such as protocols or interfaces, are relatively straightforward to identify, since their boundaries are clearly defined and easily detectable. Others may present greater challenges. Specifically, long comment blocks associated with short functions, or interleaved sequences of directives and function definitions. 
In the presence of nested structures, we select the largest enclosing entity as the chunk. For example, when one or more functions are wrapped by a compiler directive, the resulting segment is defined as a 'directive' chunk (Fig. \ref{fig:chunks-examples} E). In contrast, if a compiler directive appears within a function, the entire function is treated as a single 'function' chunk (Fig. \ref{fig:chunks-examples} D and F).
Many semantically related code elements can and should be grouped into coherent units to simplify subsequent code processing (Fig. \ref{fig:chunks-examples} B, C and F).

\textit{Fine-grained chunking.} More fine-grained chunking can be performed iteratively to extract functions from directive blocks, as well as to isolate specific class sections (public, protected or private) and namespace components.

\textit{Large files processing.} The input sequence length is limited to 8K tokens. Files exceeding this limit are processed iteratively by partitioning them into consecutive 8K-token segments. After processing the first segment, the corresponding portion is removed from the input, and the procedure is repeated on the remaining content until the entire file has been processed.

\subsection{Chunker Training Procedure}

We used Qwen2.5-Coder-32B-Instruct model (hereafter referred to as \textit{Oracle Chunker}) to generate semantic chunks with attribute associated with each chunk. The model was applied to a dataset of 39,371 files (about 73M tokens) using an instruction-based prompt augmented with a small number of illustrative examples (see Appendix 1 for the prompt). Tables \ref{tab:train-data} provide a detailed description of the data generated by Oracle Chunker for model training. 
To assess the performance of the Oracle Chunker, we first evaluated it on a manually created dataset, called the \textit{Manual Test set}.
We manually annotated several hundred code files with a few extensions. The files were randomly selected from the following repositories \cite{openssl}, \cite{redis}, \cite{bullet3},\cite{llvm}, \cite{serilog}. This manually curated dataset, although limited in size, is highly reliable. A separate part of it was also later used as a validation set, to evaluate the accuracy of our trained models.

\begin{table}[]
    \centering
    \begin{tabular}{|c|c|c|c|c|c|}
    \hline
      Type & C &  C++ & C\# & H  & Avg\\
      \hline
    Chunking accuracy (EM) & 94.9 &	87.5 & 98.8 & 78.9 & 90.03\\
    Category accuracy & 98.3 &	97.3 & 94.8 & 93.9 & 96.08\\
    \hline
    Files & 297 & 154  & 195 & 200 & total 846\\
     \hline
    \end{tabular}
    \caption{Several hundreds files were manually annotated and used to evaluate the Oracle Chunker performance.}
    \label{tab:manual-test-data}
\end{table}

The Oracle Chunker achieved high average accuracy of 90\% for chunk boundaries detection and 96\% for category labeling on the Manual Test set. This ensures that the resulting chunks are mostly correct and that the training data is of high quality.  The distribution of files and chunk categories of the Manual Test set, as well as the Oracle Chunker results, are presented in Table \ref{tab:manual-test-data}. In all experiments, we used the exact match (EM) evaluation metric. 

\textit{Training procedure.} 
We found that the following two-stage incremental learning procedure yields the best performance, likely due to the relatively small size of the models.
\begin{itemize}
    \item In the first stage, each of the four encoder models was trained as a token-level classifier, where each token was assigned to one of three classes: chunk start, chunk end, or chunk interior.

    \item In the second stage, the classification task was made more challenging by requiring the model to assign interior tokens to one of 12 predefined categories, described in Sec.\ref{subsec:chunks-taxonomy}. Accordingly, the label space was expanded to 14 classes: chunk start, chunk end, and 12 interior categories. After this stage, each token within a chunk interior was classified into one of these 12 categories, and the final category assigned to a chunk was determined based on the classification of its last token.
\end{itemize}

\textit{Training details.} We used the AdamW optimizer with a linear learning-rate scheduler and warmup. To mitigate overfitting, dropout and weight decay were applied during training. All models were trained for 5 epochs. 
In the first stage, the learning rate was set to 5e-5 for the 17M, 68M, and 150M parameter models, and to 8e-5 for the 32M model. In the second stage, a uniform learning rate of 1e-5 was used for all four models.

\subsection{Evaluation Methodology and Test Data}

We evaluated the accuracy of chunk boundaries and their associated categories using two test sets: the Manual Test set described above, and a larger \textit{Automatic Test set} generated by the Oracle Chunker using the same prompt as that employed for training set generation. 

Automatic Test set contains nearly 1,000 examples of each of C, C++, C\#, H, M/MM files selected mostly from the repositories \cite{openssl}, \cite{redis}, \cite{bullet3},\cite{llvm}, \cite{serilog}, and processed by Oracle Chunker with corresponding ground truth attributes. We compared the Oracle chunks with those produced by our method. Detailed results for chunking performance and category classification accuracy are presented in Table \ref{tab:automatic-test-data-chunks}.
The results demonstrate that, despite the model small size of up to 150M parameters, they are capable of achieving strong semantic code understanding, with performance comparable to that of much larger models. This is in particular evident for the smallest model with only 17M parameters.

We also evaluated a larger language model from the same family as our Oracle chunker, specifically a Qwen-Coder-7B model, using the same prompt configuration as in the Oracle setting. We compared the accuracy of chunk boundary detection and category assignment produced by our lightweight classification models against this substantially larger LLM (see Table \ref{tab:automatic-test-data-chunks}). The results indicate that employing a larger model, even of moderate scale such as 7B parameters, is not always justified. In many cases, a properly trained 17M-parameter model matches or even outperforms the 7B model. Moreover, when averaging results across programming languages, our smaller classifiers demonstrate strong capability in capturing programming language semantics and structural relationships. 

Notice that the ability of general purpose SLMs to follow complex instructions in code understanding tasks may degrade substantially as model size decreases. For example, the Qwen-0.5B model (with the same prompt as for its bigger counterparts) exhibits extremely low accuracy on the Automatic test set, achieving on average less than 15\% both for chunking and for category classification. We do not report the detailed results for the Qwen 0.5B model due to their limited relevance.

\begin{table}[]
    \centering
    \scriptsize
    \setlength{\tabcolsep}{2pt}
    \begin{tabular}{|c|c|c|c|c|c|}
    \hline
       & C & C++ & C\# & H & Avg \\
       \hline
     Model & 
    \begin{tabular}{c|c} EM & Acc \end{tabular}& 
    \begin{tabular}{c|c} EM & Acc \end{tabular}&
    \begin{tabular}{c|c} EM & Acc \end{tabular}& 
    \begin{tabular}{c|c} EM & Acc \end{tabular}&
    \begin{tabular}{c|c} EM & Acc \end{tabular}\\
      \hline
    17M & 
    \begin{tabular}{c|c}85.6 & 95.5\end{tabular} & 
    \begin{tabular}{c|c}86.8 & 95.2\end{tabular} & 
    \begin{tabular}{c|c}94.3 & 93.9\end{tabular} & 
    \begin{tabular}{c|c}63.4 & 92.3\end{tabular}&
    \begin{tabular}{c|c}\textbf{82.5} & \textbf{94.2}\end{tabular}\\
    32M & 
    \begin{tabular}{c|c}88.9 & 96.8\end{tabular}& 
    \begin{tabular}{c|c}91.2 & 95.8 \end{tabular}& 
    \begin{tabular}{c|c}95.8 & 96.6\end{tabular}& 
    \begin{tabular}{c|c}65.1 & 92.4 \end{tabular}&
    \begin{tabular}{c|c}\textbf{85.3} & \textbf{95.4}\end{tabular}\\
    68M & 
    \begin{tabular}{c|c}90.0 & 96.3 \end{tabular}& 
    \begin{tabular}{c|c}92.3 & 91.9 \end{tabular}& 
    \begin{tabular}{c|c}96.5 & 93.7\end{tabular} &  
    \begin{tabular}{c|c}65.9 & 93.2\end{tabular}&
    \begin{tabular}{c|c}\textbf{86.2} & \textbf{93.8}\end{tabular}\\
    150M & 
    \begin{tabular}{c|c}89.6 & 96.4 \end{tabular}& 
    \begin{tabular}{c|c}94.9 & 91.5 \end{tabular}& 
    \begin{tabular}{c|c}83.6 & 86.8 \end{tabular}& 
    \begin{tabular}{c|c}96.5 & 98.6\end{tabular}&
    \begin{tabular}{c|c}\textbf{91.2} & \textbf{93.3}\end{tabular}\\
     \hline
     \tiny Oracle Chunker & 
    \begin{tabular}{c|c}94.9 & 98.3 \end{tabular}& 
    \begin{tabular}{c|c}87.5 & 97.3 \end{tabular}& 
    \begin{tabular}{c|c}98.8 & 94.8 \end{tabular}& 
    \begin{tabular}{c|c}78.9 & 93.9 \end{tabular}&
    \begin{tabular}{c|c}90.0 & 96.1\end{tabular}\\
     \hline
    \end{tabular}
    \caption{The evaluation results of our four models on the \textit{Manual Test set} described in Table \ref{tab:manual-test-data}. For each PL and model we report two metrics: chunking accuracy (EM) and category classification accuracy (Acc). In the bottom row. we include the results of the Oracle Chunker, previously presented in Table \ref{tab:manual-test-data}, for ease of comparison.}
    \label{tab:manual-test-data-ettins}
\end{table}

\begin{table}[]
    \centering
    \scriptsize
    \setlength{\tabcolsep}{2pt}
    \begin{tabular}{|c|c|c|c|c|c|c|}
    \hline
       & C & C++ & C\# & H  & M/MM & Avg\\
      \hline
    Model  & 
    \begin{tabular}{c|c} EM & Acc \end{tabular}& 
    \begin{tabular}{c|c} EM & Acc \end{tabular}&
    \begin{tabular}{c|c} EM & Acc \end{tabular}& 
    \begin{tabular}{c|c} EM & Acc \end{tabular}& 
    \begin{tabular}{c|c} EM & Acc \end{tabular}& 
    \begin{tabular}{c|c} EM & Acc \end{tabular}\\
    \hline
    17M & 
    \begin{tabular}{c|c} 76.3 & 90.7 \end{tabular}& 
    \begin{tabular}{c|c} 78.6 & 92.6 \end{tabular}&
    \begin{tabular}{c|c} 82.1 & 89.6 \end{tabular}& 
    \begin{tabular}{c|c} 75.5 & 93.0 \end{tabular}& 
    \begin{tabular}{c|c} 68.4 & 81.8 \end{tabular}& 
    \begin{tabular}{c|c} \textbf{76.2} & \textbf{89.5} \end{tabular}\\
    32M & 
    \begin{tabular}{c|c} 80.1 & 89.6 \end{tabular}& 
    \begin{tabular}{c|c} 79.6 & 83.2 \end{tabular}& 
    \begin{tabular}{c|c} 82.4 & 61.9 \end{tabular}&
    \begin{tabular}{c|c} 79.7 & 84.9 \end{tabular}& 
    \begin{tabular}{c|c} 76.1 & 70.9 \end{tabular}& 
    \begin{tabular}{c|c} \textbf{79.6} & \textbf{78.1} \end{tabular}\\
    68M & 
    \begin{tabular}{c|c} 81.0 & 91.7 \end{tabular}&
    \begin{tabular}{c|c} 77.8 & 91.3 \end{tabular}& 
    \begin{tabular}{c|c} 82.7 & 66.5 \end{tabular}& 
    \begin{tabular}{c|c} 79.4 & 90.1 \end{tabular}& 
    \begin{tabular}{c|c} 76.2 & 80.4 \end{tabular}& 
    \begin{tabular}{c|c} \textbf{79.4} & \textbf{84.0} \end{tabular}\\
    150M & 
    \begin{tabular}{c|c} 80.9 & 93.8 \end{tabular}&
    \begin{tabular}{c|c} 79.4 & 90.6 \end{tabular}& 
    \begin{tabular}{c|c} 88.4 & 96.3 \end{tabular}& 
    \begin{tabular}{c|c} 80.2 & 94.1 \end{tabular}& 
    \begin{tabular}{c|c} 77.7 & 88.6 \end{tabular}& 
    \begin{tabular}{c|c} \textbf{81.3} & \textbf{92.7} \end{tabular}\\ 
    Qwen7B & 
    \begin{tabular}{c|c} 82.1 & 92.8 \end{tabular}&
    \begin{tabular}{c|c} 81.1 & 93.0 \end{tabular}&
    \begin{tabular}{c|c} 66.1 & 82.4 \end{tabular}& 
    \begin{tabular}{c|c} 83.6 & 74.5 \end{tabular}& 
    \begin{tabular}{c|c} 79.2 & 87.8 \end{tabular}& 
    \begin{tabular}{c|c} 78.4 & 86.1 \end{tabular}\\
    \hline
    Chunks & 7068 & 6228 & 2848 & 8155 & 6936 & \tiny total 31,235 \\
     \hline
    \end{tabular}
    \caption{The evaluation results of our four chunking models and Qwen-Coder-7B on the Automatic Test set. For each PL and model we report two metrics: chunking accuracy (EM) and category classification accuracy (Acc).}
    \label{tab:automatic-test-data-chunks}
\end{table}

\subsection{Pretraining Justification and Impact}
\label{subsec:justify-pretrain}

To justify encoder pretraining on diverse code data (Section \ref{subsec:FT}), we trained an additional family of chunker models built on the original Ettin encoders. We then compared the accuracy of chunkers initialized by the original encoders with those initialized by encoders after the fine-tuning stage described in Section \ref{subsec:FT}. Table \ref{tab:automatic-test-no-pretraining} reports the chunking accuracy (EM) on the Automatic test set for models trained based on the original encoders, along with the corresponding accuracy gains obtained when the chunkers are initialized by the fine-tuned encoders. As shown in the table, incorporating code knowledge through fully unsupervised pretraining improves chunking performance in nearly all cases (with one exception), yielding an average gain of 3-6\% across models and PLs. The greatest impact of pretraining is observed for the smallest (17M) model and for C\# language.

\begin{table}[]
    \centering
    \scriptsize
    \setlength{\tabcolsep}{2pt}
    \begin{tabular}{|c|c|c|c|c|c|c|}
    \hline
      Model & C & C++ & C\# & H  & M/MM & Avg\\
      \hline
     17M &
      73.4 \textbf{(+2.9)} &
      77.2 \textbf{(+1.4)} & 
      57.2 \textbf{(+24.9)} &
      74.2 \textbf{(+1.3)} &
      66.1 \textbf{(+2.3)} &
      69.6 \textbf{(+6.6)}\\
    32M &
    77.8 \textbf{(+2.3)} &
    78.4 \textbf{(+1.2)} & 
    60.3 \textbf{(+22.1)} &
    76.9 \textbf{(+2.8)} &
    73.7 \textbf{(+2.4)} &
    73.4 \textbf{(+6.2)}\\
    68M &
    80.2 \textbf{(+0.8)} &
    78.1 (-0.3) & 
    68.6 \textbf{(+14.1)} &
    78.1 \textbf{(+1.3)} &
    75.8 \textbf{(+0.4)} &
    76.4 \textbf{(+3.0)}\\
    150M &
    80.5 \textbf{(+0.4)} &
    78.7 \textbf{(+0.7)} & 
    60.0 \textbf{(+28.4)} &
    77.9 \textbf{(+2.3)} &
    76.9 \textbf{(+0.8)} &
    74.8 \textbf{(+6.5)}\\ 
    \hline
    \end{tabular}
    \caption{The chunking accuracy (EM) on the Automatic test set for models trained on top of the original Ettin encoders (+ the corresponding accuracy gains obtained when the chunkers are initialized by the fine-tuned encoders, Sec. \ref{subsec:FT}).
    }
    \label{tab:automatic-test-no-pretraining}
\end{table}

\section{Experiments and results}
\label{sec:experiments} 

In the previous section, we argued that our chunker models are compact yet accurate. In this section, we evaluate the impact of our accurate chunking on downstream repository-related tasks. We consider two tasks - retrieval and code generation, and evaluate them using publicly available datasets.

\subsection{Retrieval Task}

We evaluated the ability of our chunking methods to support repository-level retrieval tasks using the RepoQA benchmark. The authors built a Searching Needle Function (SNF) task that challenges models to locate functions based on their natural-language descriptions.

In our experiments, we selected ten of the fifty repositories included in the benchmark, focusing specifically on programming languages from the C family. Each repository was preprocessed by splitting files into chunks and embedding each chunk using the modern-gte embedder \cite{gte-embedder}. Each code snippet corresponding to a “needle” function in RepoQA is associated with two types of natural language annotations: a general \textit{description} and a shorter \textit{purpose}. The retrieval task consisted of identifying the correct code snippet by finding the embedding vector most similar to the embedding of the description and the purpose, separately. 
Table\ref{tab:repoqa} reports the total number of chunks generated by each method, along with the retrieval performance, measured by the nDCG@10 metric. We compare the performance of our four models of different sizes with the standard Tree-sitter chunker. As shown in the table, our approach achieves comparable or superior retrieval accuracy while producing substantially fewer chunks, thereby significantly accelerating vector database search. Also, restricting retrieval to chunks of specific types (e.g., 'function') or excluding certain categories (e.g., 'directive' or 'comment') improves performance while reducing computational cost.

\begin{table}[]
\scriptsize
    \centering
    \setlength{\tabcolsep}{4pt}
    \begin{tabular}{|c|c|c|c|c|c|}
    \hline
      Input & TS  & 17M & 32M & 68M & 150M \\
      \hline
    Description & 75.81 \% & 75.49\% & 74.57\% & \textbf{76.11}\% & \textbf{78.01}\% \\
    Purpose & 66.37\% & \textbf{66.39}\% & 64.41\%  & \textbf{67.18}\% & \textbf{67.35}\% \\
     \hline
     Chunks & 21,195 & 13,925 & 14,977 & 13,999 & 15,151 \\ 
     \hline
    \end{tabular}
    \caption{The table reports the retrieval performance, measured using the nDCG@10 metric, for our four models in comparison with the Tree-sitter chunker. The bottom row shows the total number of chunks generated by each method.}
    \label{tab:repoqa}
\end{table}

\subsection{Code Generation Task}

YABloCo benchmark \cite{yabloco}  targets long-context code generation at the function level, particularly for C and C++ - languages, and context that many previous benchmarks did not cover. The benchmark consists of over 200 tasks, drawn from large repositories: \cite{openssl}, \cite{redis}, \cite{bullet3}, \cite{llvm}. Models are evaluated by their ability to generate correct function body given extensive surrounding context. 

We evaluate the code generation accuracy of Qwen2.5-Coder-32B-Instruct model under five settings: (1) without context, (2) with context provided by the Tree-sitter-based chunking tool, (3) with context generated by an LLM (Qwen2.5-Coder-32B-Instruct), (4) with context produced by our four chunker models, and (5) with the 'oracle' context provided by the \cite{yabloco}. In the original YABloCo paper, the 'oracle' context was defined by the authors as all the functions in the repository along the docstrings and function bodies that are invoked into the target function body. For the settings (2)-(4), repository files were first segmented into chunks using the corresponding methods. These chunks were then embedded using the gte-modernbert embedder \cite{gte-embedder}, after which the top 10 most relevant chunks were retrieved and supplied as context to the code generation model. We evaluated code generation performance by computing pass@1 and pass@10 metrics across the YABLoCo benchmark repositories (see Table \ref{tab:yabloco-generations}).

\begin{table}[]
\scriptsize
\centering
\setlength{\tabcolsep}{1pt}
\begin{tabular}{cc}
a) &
\setlength{\tabcolsep}{3pt}
\begin{tabular}{|c|c|c|c|c|c|}
\hline
repo & no content & TS & LLM  & 17M / 32M / 68M / 150M & oracle\\
\hline
openssl & 18.33 & 26.67 & 28.33  & \textbf{30} / \textbf{30} / \textbf{30} / \textbf{28.33} & 30\\
redis & 14.93 & 16.42 & 22.39  & 20.9 / \textbf{22.39} / \textbf{22.39} / \textbf{25.37}  & 23.88\\
bullet3 & 39.13& 39.13& 43.48  & \textbf{47.83} / 34.78 / \textbf{39.13} / 34.78 & 73.91\\
llvm & 13.24& 14.71& 13.24  & \textbf{14.71} / \textbf{14.71} / 13.24 / 13.24 & 26.47\\
\hline
\end{tabular} \\ \\
b) &
\setlength{\tabcolsep}{3pt}
\begin{tabular}{|c|c|c|c|c|c|}
\hline
    repo &  no content & TS & LLM  & 17M / 32M / 68M / 150M & oracle\\
\hline
openssl &  30 & 35 & 41.67  & \textbf{45} / \textbf{45} / 40 / 40 & 45\\
redis &  17.91 & 28.36 & 28.36  & \textbf{28.36} / \textbf{31.34} / \textbf{29.85} / \textbf{28.36} & 29.85\\
bullet3 &  47.83& 52.17& 56.52& 52.17 / 43.48 /52.17 / 47.83& 82.61\\
llvm &  17.56 & 19.12 & 19.12 & \textbf{22.06} / \textbf{26.47} / \textbf{23.53} / \textbf{20.59} & 29.41\\
\hline
\end{tabular}

\end{tabular}
\caption{Code generation results for the Qwen2.5-Coder-32B-Instruct model for different context inputs: a) single generation, temperature = 0; b) 10 generations.}
\label{tab:yabloco-generations}
\end{table}

As evident from the Table \ref{tab:yabloco-generations}, the context provided by our chunkers substantially improves code generation accuracy compared to the 'no context' setting, and in most cases outperforms the context produced by the tree-sitter and the large model. In some instances, it even surpasses performance achieved using the oracle context. Notably, larger chunker models do not yield consistently better results; in several cases, the 17M model provides sufficiently high-quality chunks. This suggests that more efficient, relatively small models are very effective for repository processing in a range of downstream tasks.

\subsection{Inference}

The inference times on both GPU and CPU, for a batch size of 1 and two context lengths (512 and 8K tokens), are reported in Table \ref{tab:inference-time}. The reported results reflect model inference only and exclude both tokenization and postprocessing. Tokenization process introduces an additional latency of approximately 0.1–0.2 seconds, with substantial variability across iterations. Furthermore, the GPU used in this study does not support FlashAttention, resulting in comparatively slower performance.  The analysis of inference times suggests that at least the two smallest models (17M and 32M parameters) are suitable for execution on a CPU, particularly in scenarios involving minor, localized code changes.

\section{Conclusion}
\label{sec:conclusion}

This paper presents SemChunk-C - a new lightweight approach to semantic chunking of code written in C-family programming languages. By training the models of various sizes (from 17M to 150M parameters) we investigate the minimal model size capable of capturing semantic nuances in code, and accurately aggregating lines of code based on contextual requirements. In addition, we examine the role of repository-level chunking and demonstrate its importance for improving performance in downstream tasks.
\begin{table}[]
\scriptsize
\centering
\setlength{\tabcolsep}{3pt}
\begin{tabular}{|c|c|c|c|c|}
    \hline
    & GPU (\tiny{context = 512}) & GPU (\tiny{context = 8K}) & CPU (\tiny{context = 512})  & CPU (\tiny{context = 8K})\\
    \hline
    17M  &  0.00878 & 0.0878 & 0.03439 &  0.695\\
    32M  &  0.01393 & 0.177 & 0.0572 &  1.158 \\
    68M  &  0.0217 &  0.476 & 0.1204 &  2.609 \\
    150M  & 0.0242 & 0.84 & 0.205 & 4.4699 \\
    \hline
    \end{tabular}
    \caption{GPU and CPU inference times (seconds) with batch size = 1.}
    \label{tab:inference-time}
\end{table}
\appendix
\section{Appendix}

Below is a prompt we used for Qwen32B-Coder-Instruct to generate code chunks and their corresponding attributes.

\footnotesize
"""
Instruct: You are provided with a code snippet.
Analyze this code and split it into blocks. Do not generate new text or code, just add delimiters and block attribute. Add delimiter 'block start' at the beginning of the block and delimiter 'block end' at the end of the block. Blocks may be of the following types.

A comment: if a comment is standalone (like starting comments in the file). If comments precede the function, put the comments and the function into a single block.
If it's a comment, add 'attribute: comment' after 'block end'.

A function: it may be a free, member or inline function, template, constructor or destructor, lambda definition, operator overloads, record, extern for functions. If a function or struct contains sub-functions or compiler directives inside, don't split it, put into a single block.
If it's a function, add 'attribute: function' after 'block end'.

A class: if a code relates to a class or method.
If it's a class, add 'attribute: class' after 'block end'.
A type definition block: if it's a typedef, union, struct, enum or alias.
If it's a type function, add 'attribute: type\_definition' after 'block end'.

A namespace: if it's a namespace.
If it's a namespace, add 'attribute: namespace' after 'block end'.

An include: if the code is \#include, using, import, @import or export.
If it's an include, add 'attribute: include' after 'block end'.

A compiler directive: if the row starts with \# (like \#ifdef, \#error etc), but is not an include. Put everything between \#if and corresponding \#endif into a single block, even if there are functions, directives or includes inside it.
If it's a directive, add 'attribute: directive' after 'block end'.

An implementation: if it starts from @implementation and ends with @end. In this case add 'block end' only after the corresponding @end.
If it's an implementation, add 'attribute: implementation' after 'block end'.

A protocol: if it starts from @protocol and ends with @end. In this case add 'block end' only after the corresponding @end.
If it's a protocol, add 'attribute: protocol' after 'block end'.

An interface: if it starts from @interface and ends with @end. In this case add 'block end' only after the corresponding @end.
If it's an interface, add 'attribute: interface' after 'block end'.

Variables: if the code defines global or local variables, delegate or extern for variables.
If it's a variable, add 'attribute: variable' after 'block end'.

If a code snippet is not one of the above types, also wrap it by 'block start' and 'block end', and add 'attribute: misc' after 'block end'.
Don't explain and do not write what is the purpose of the block.

Follow the example below: \{\}

Code snippet: \{\}
"""

\bibliographystyle{unsrt}  
\bibliography{references} 

\end{document}